\documentstyle[preprint,aps]{revtex}
\tighten
\begin{document}
\draft
\preprint{\vbox{Submitted to Physics Letters B \hfill FSU-SCRI-95-67 \\
                                          \null\hfill nucl-th/9507030}}
\title{Isospin Violations in the Pion-Nucleon System}
\author{J. Piekarewicz}
\address{Supercomputer Computations Research Institute, \\
         Florida State University, Tallahassee, FL 32306}
\date{\today}
\maketitle

\begin{abstract}
We examine the effect of isospin-violating meson-nucleon coupling
constants on low-energy pion-nucleon scattering. We compute the
couplings in the context of a nonrelativistic quark model. The
difference between the up and down constituent masses induces a
coupling of the neutral pion to the proton that is slightly larger
than the corresponding one for the neutron. This difference
generates a large isospin-violating correction---proportional
to the isospin-even contribution arising from the nucleon Born
terms---to the charge-exchange ($\pi^{-}p \rightarrow \pi^{0}n$)
amplitude. In contrast to the isospin-conserving case, this
correction is not cancelled by $\sigma$-meson exchange; in our
model there is no isospin-violating $NN\sigma$ coupling at
$q^2=0$. As a result, we find a violation of the triangle
identity consistent with the one reported by Gibbs, Ai, and
Kaufmann from a recent analysis of pion-nucleon data.
\end{abstract}
\pacs{PACS number(s):~11.30.-j, 13.75.Gx}

\narrowtext

\section{Introduction}
\label{secintro}

  Low energy pion-nucleon ($\pi N$) scattering is one of the best
available tools for testing small violations to approximate
symmetries of nature. Such violations are expected to be amplified
in low-energy $\pi N$ scattering because of the constraints imposed
on the symmetry-conserving amplitudes by chiral symmetry. At low
energies (i.e., in the soft-pion limit) the pions couple very
weakly to the nucleons as a direct consequence of chiral symmetry.
Thus, although the violations to the symmetry might be small,
they must be considered relative to intrinsically small
symmetry-conserving amplitudes.

  An example of such a scenario has been reported recently by
Gibbs, Ai, and Kaufmann~\cite{gak95}. They have analyzed
low-energy pion-nucleon data in search of isospin violations.
{}From very precise data on elastic ($\pi^{\pm}p$) and
charge-exchange ($\pi^{-}p \rightarrow \pi^{0}n$) reactions they
have extracted $\pi N$ scattering amplitudes from which they
have computed violations to the ``triangle identity''
\begin{equation}
   D \equiv f(\pi^{-}p \rightarrow \pi^{0}n) -
  {1 \over \sqrt{2}} \, \Big[f(\pi^{+}p)-f(\pi^{-}p)\Big] \;.
 \label{triangle}
\end{equation}
They observed a large isospin violation---of the order of 7\%---even
after accounting for Coulomb effects and hadronic mass differences.
This is particularly interesting since isospin-breaking mechanisms,
having their origin in the up-down quark mass difference and
electromagnetic effects, are expected to be present at
the $\sim$~1\% level.

  Evidence for the loss of isospin symmetry in the nucleon-nucleon
($NN$) system is well documented. The difference in the $pp$ and $nn$
scattering lengths~\cite{slaus89}, the Nolen-Schiffer
anomaly\cite{nolsch69,miller90}, and the neutron-proton
analyzing-power difference~\cite{knut90,abegg89,abegg94}
are all well known examples. Most theoretical efforts directed at
understanding isospin-violating observables in the $NN$ system proceed
from a two-body interaction constrained from fits to two-nucleon data
and incorporate isospin-violating corrections from a variety of sources.
These can be classified as arising from:
(i) isovector-isoscalar mixing in the meson propagator---such as
    $\rho$-$\omega$ mixing,
(ii) isospin-breaking in the nucleon wave function---through the
      neutron-proton mass difference, and
(iii) isospin-breaking in the meson-nucleon and photon-nucleon
      vertices---as in the case of electromagnetic scattering.
It is important to note that all these isospin-breaking
mechanisms also operate in the pion-nucleon system. Thus,
a clear understanding of their role in $NN$ scattering could
be of great value to the analysis of low-energy $\pi N$ data.
A particularly important---and timely---example
is $\rho$-$\omega$ mixing. Naively, one would expect
large violations to the triangle identity (also known as the
``triangle discrepancy'') to arise from $\rho$-$\omega$ mixing
because of the strong $NN\omega$ and $\pi\pi\rho$ couplings.
Note, however, that in computing near-threshold $\pi N$
observables it is the mixing amplitude near $q^2=0$ that is
relevant. The traditional mechanism of $\rho$-$\omega$ mixing, with
the mixing amplitude fixed at the on-shell point, has been called
recently into question~\cite{ght92}. Indeed, a large number of
calculations using a variety of models have found a value of the
$\rho$-$\omega$ mixing amplitude at $q^2=0$ that is strongly
suppressed relative to its on-shell
value~\cite{piewil93,hats93,krein93,mitch94,oconn94}.
Moreover, for models in which the vector mesons couple to conserved
currents, the $\rho$-$\omega$ mixing amplitude is identically
zero at $q^2=0$~\cite{piewil93,oconn94}. Thus, we believe that
$\rho$-$\omega$ mixing should play a small role in low-energy
pion-nucleon scattering.

  Removing $\rho$-$\omega$ mixing as a viable source of isospin-breaking
has important phenomenological consequences; on-shell $\rho$-$\omega$
mixing accounts for a substantial fraction of the neutron-proton
analyzing-power difference at 183 MeV~\cite{knut90,miller86,willia87}.
Hence, if $\rho$-$\omega$ mixing is no longer important at
$q^2$~{\raise2pt\hbox to 8pt{\raise -6pt\hbox{$\sim$}\hss{$<$}}}~0,
additional sources of isospin violation must be found. In a recent study
of hadronic structure Dmitra\v{s}inovi\'c and Pollock have computed
isospin-violating corrections to the electroweak form factors of the
nucleon~\cite{dmitra95}. Motivated by their findings we have investigated
new sources of charge-symmetry violation in the $NN$ potential which
resulted from isospin-violating meson-nucleon coupling
constants~\cite{ghp95}. The resulting class IV contribution to the
charge-symmetry-breaking $NN$ potential is comparable in magnitude and
identical in sign to the one obtained from on-shell $\rho$-$\omega$
mixing. We showed that this new contribution---without on-shell
$\rho$-$\omega$ mixing---is consistent with the measured value of
$\Delta A$ at 183 MeV~\cite{ghp295}. It is the purpose of this paper
to estimate the effect of isospin-violating meson-nucleon coupling
constants on low-energy pion-nucleon scattering.

\section{Low-energy pion-nucleon scattering}
\label{secpin}

  We approach the study of low-energy pion-nucleon scattering in a
conventional way; we include contributions arising from the (s- and
u-channel) nucleon Born terms and from (t-channel) meson
exchanges~\cite{camp78}. These contributions---particle-exchange
poles---give a good representation of the amplitude when the
poles are close to the physical region, such as in low-energy $\pi N$
scattering in the chiral ($m_{\pi} \rightarrow 0$) limit. The linear
$\sigma$-model~\cite{gell60} and Quantum Hadrodynamics
(QHD-II)~\cite{serwal86} are appropriate theoretical frameworks to
generate these tree-level contributions. The models differ, at tree level,
in the allowed t-channel exchanges and, hence, in the prediction of
low-energy $\pi N$ parameters. However, as we shall see, they generate
the same isospin-violating contributions in our model.

  The $\pi N$ scattering matrix can be written in terms of two
sets (one for each isospin combination) of two Lorentz invariant
amplitudes ($A$ and $B$) which contain all dynamical information
about the reaction~\cite{camp78}
\begin{equation}
  \hat{{\cal T}} =
   \Big[ A^{(+)}(s,t)+{1 \over 2}({\rlap/k}+{\rlap/k'})B^{(+)}(s,t)\Big] -
   \Big[ A^{(-)}(s,t)+{1 \over 2}({\rlap/k}+{\rlap/k'})B^{(-)}(s,t)\Big]
   ({\bf T}\cdot\mbox{\boldmath$\tau$}) \;.
 \label{relt}
\end{equation}
Note, the Lorentz invariant amplitudes are written in terms of the
relevant Mandelstam variables ($t \equiv q^{2}$)
\begin{mathletters}
 \begin{eqnarray}
   s &=& (p+k)^{2}=(p'+k')^{2} \;, \\
   t &=& (k-k')^{2}=(p'-p)^{2} \;, \\
   u &=& (p-k')^{2}=(p'-k)^{2} \;,
 \label{mandel}
 \end{eqnarray}
\end{mathletters}
where $k(k')$ and $p(p')$ are the initial(final) four-momenta of the
pion and nucleon, respectively. The Mandelstam variables
are related by $s+t+u=2m_{\pi}^{2}+2M^{2}$. We have also introduced
pion (${\bf T}$) and nucleon (\mbox{\boldmath$\tau$}) isospin matrices
[note, $(T_{a})_{bc} \equiv -i\epsilon_{abc}$]. Isospin invariance,
which is still assumed unbroken, allows for only two isospin
combinations: isospin even [denoted by $(+)$] and isospin odd
[denoted by $(-)$]. The connection to the reaction amplitudes is
given through the following relations:
\begin{mathletters}
 \begin{eqnarray}
   {\cal T}(\pi^{+}p \rightarrow \pi^{+}p) &=&
   {\cal T}^{(+)} - {\cal T}^{(-)}           \;,  \\
   {\cal T}(\pi^{-}p \rightarrow \pi^{-}p) &=&
   {\cal T}^{(+)} + {\cal T}^{(-)}           \;,  \\
   {\cal T}(\pi^{-}p \rightarrow \pi^{0}n) &=&
        - \sqrt{2}\,{\cal T}^{(-)}                \;.
  \label{reacmpl}
 \end{eqnarray}
\end{mathletters}
{}From these, the triangle identity [see Eq.~(\ref{triangle})] follows
by inspection.

  The partial-wave decomposition of the scattering amplitude is
simplest if carried out after the Lorentz-invariant scattering
matrix has been evaluated between on-shell spinors in the
center-of-mass (CM) frame. Thus, as an operator in the spin
space of the nucleon the $\pi N$ scattering amplitude can be
written as,
\begin{equation}
  \hat{f}^{(\pm)} =
   f^{(\pm)}_{1}(W,\theta) + f^{(\pm)}_{2}(W,\theta)
   {(\mbox{\boldmath$\sigma$}\cdot{\bf k}')
    (\mbox{\boldmath$\sigma$}\cdot{\bf k} ) \over k^2} \;,
  \label{pinamp}
\end{equation}
where the connection to the Lorentz-invariant amplitudes is
given through the relations
\begin{mathletters}
 \begin{eqnarray}
   f^{(\pm)}_{1}(W,\theta) &=&
     \left({E_{k}+M \over 8\pi W}\right)
     \Big[ +A^{(\pm)}(s,t) + (W-M)B^{(\pm)}(s,t) \Big] \;, \\
   f^{(\pm)}_{2}(W,\theta) &=&
     \left({E_{k}-M \over 8\pi W}\right)
     \Big[ -A^{(\pm)}(s,t) + (W+M)B^{(\pm)}(s,t) \Big] \;.
 \label{f12}
 \end{eqnarray}
\end{mathletters}
Here $\theta$ denotes the CM scattering angle and $W=(\epsilon_{k}+E_{k})$
is the total energy of the system in the CM frame; it is written in terms
of the individual pion ($\epsilon_{k}$) and nucleon ($E_{k}$) contributions.
Finally, by introducing the partial-wave amplitudes
$f{\lower 2pt \hbox{$\scriptstyle l^{\pm}$}}$, appropriate
for scattering in a total angular-momentum channel $j=l\pm 1/2$, the
amplitudes $f_{1}$ and $f_{2}$ can be expanded in a partial-wave series:
\begin{mathletters}
 \begin{eqnarray}
   f^{(\pm)}_{1}(W,\theta) &=& \sum_{l}
     \Big[
       f^{(\pm)}_{\lower 2pt \hbox{$\scriptstyle l^{+}$}}(W)
       P^{'}_{l+1}(\cos\theta) -
       f^{(\pm)}_{\lower 2pt \hbox{$\scriptstyle l^{-}$}}(W)
       P^{'}_{l-1}(\cos\theta)
     \Big]  \;, \\
   f^{(\pm)}_{2}(W,\theta) &=& \sum_{l}
     \Big[
       f^{(\pm)}_{\lower 2pt \hbox{$\scriptstyle l^{-}$}}(W) -
       f^{(\pm)}_{\lower 2pt \hbox{$\scriptstyle l^{+}$}}(W)
     \Big] P^{'}_{l}(\cos\theta) \;.
 \label{f12pw}
 \end{eqnarray}
\end{mathletters}

We compute the Lorentz invariant amplitudes $A$ and $B$ in the linear
sigma model~\cite{camp78}. The connection to other models, specifically
to QHD, will be done below. At tree-level, the amplitudes receive
contribution from only three Feynman diagrams: the two nucleon Born
terms and $\sigma$-meson exchange. That is,
\begin{mathletters}
 \begin{eqnarray}
  A^{(+)}(s,t) &=&
   -{g_{\lower 2pt \hbox{$\scriptstyle NN\pi$}}^{2}\over M}
    {m_{\sigma}^{2}-m_{\pi}^{2} \over t-m_{\sigma}^{2}}
    \mathop{\longrightarrow}\limits_{|{\bf k}| \rightarrow 0} \;
    {g_{\lower 2pt \hbox{$\scriptstyle NN\pi$}}^{2}\over M}
    \left(1-{m_{\pi}^{2} \over m_{\sigma}^{2}}\right)         \; \\
  A^{(-)}(s,t) &=& 0 \;, \\
  B^{(+)}(s,t) &=&
   -{g_{\lower 2pt \hbox{$\scriptstyle NN\pi$}}^{2} \over s-M^{2}}
   +{g_{\lower 2pt \hbox{$\scriptstyle NN\pi$}}^{2} \over u-M^{2}}
    \mathop{\longrightarrow}\limits_{|{\bf k}| \rightarrow 0} \;
   -{g_{\lower 2pt \hbox{$\scriptstyle NN\pi$}}^{2}\over Mm_{\pi}}
    \left(1-{m_{\pi}^{2} \over 4M^{2}}\right)^{-1}            \; \\
  B^{(-)}(s,t) &=&
   -{g_{\lower 2pt \hbox{$\scriptstyle NN\pi$}}^{2}  \over s-M^{2}}
   -{g_{\lower 2pt \hbox{$\scriptstyle NN\pi$}}^{2}  \over u-M^{2}}
    \mathop{\longrightarrow}\limits_{|{\bf k}| \rightarrow 0}    \;
    \hskip 12pt
    {g_{\lower 2pt \hbox{$\scriptstyle NN\pi$}}^{2}\over 2M^{2}}
    \left(1-{m_{\pi}^{2} \over 4M^{2}}\right)^{-1}               \;.
 \label{bminus}
 \end{eqnarray}
\end{mathletters}
where the limit follows from evaluating the amplitudes at
threshold: $t=0$, $s=(M+m_{\pi})^{2}$, and $u=(M-m_{\pi})^{2}$.
The extraction of the $\pi N$ scattering lengths, defined by
\begin{equation}
 a_{0}^{(\pm)}=\lim_{|{\bf k}|\rightarrow 0}f^{(\pm)}_{1}
              ={1 \over 4\pi(1+m_{\pi}/M)}
               \Big[A^{(\pm)} + m_{\pi} B^{(\pm)} \Big] \;,
 \label{scattl}
\end{equation}
is now straightforward. We obtain,
\begin{mathletters}
 \begin{eqnarray}
 a_{0}^{(+)} &=& {1 \over 4\pi(1+m_{\pi}/M)}
    {g_{\lower 2pt \hbox{$\scriptstyle NN\pi$}}^{2}\over M}
    \left[\left(1-{m_{\pi}^{2} \over m_{\sigma}^{2}}\right) -
          \left(1-{m_{\pi}^{2} \over 4M^{2}}\right)^{-1}\right]
    \mathop{\longrightarrow}\limits_{m_{\pi} \rightarrow 0} 0 \;, \\
 a_{0}^{(-)} &=& {1 \over 4\pi(1+m_{\pi}/M)}
    {g_{\lower 2pt \hbox{$\scriptstyle NN\pi$}}^{2}\over M}
    \left({m_{\pi} \over 2M}\right)
    \left(1-{m_{\pi}^{2} \over 4M^{2}}\right)^{-1}
    \mathop{\longrightarrow}\limits_{m_{\pi} \rightarrow 0} 0 \;.
  \label{slengths}
 \end{eqnarray}
\end{mathletters}
The $\sigma$-exchange contribution is a direct consequence of
the underlying chiral symmetry of the model; it is essential
for effecting the sensitive cancellation of the isospin-even
scattering length. Indeed, each individual contribution to
$a_{0}^{(+)}$ is approximately two orders of magnitude larger
than the experimental value. Instead, the isospin-odd scattering
length vanishes in the chiral limit without the need for sensitive
cancellations; in the linear $\sigma$ model no additional t-channel
exchanges are included.

A model that allows for additional t-channel exchanges is
QHD-II~\cite{serwal86}. Note, even though QHD-II is not a
chiral model, a reasonable description of low-energy $\pi N$
scattering has been achieved through a ``fine tuning'' of
parameters~\cite{serwal86,matser82}. A potentially important
(t-channel) isospin-breaking contribution to $\pi N$ scattering
might come via $\rho$-meson exchange. Indeed, recently we have
computed a large isospin violation in the $NN\rho$ coupling
constant~\cite{ghp95}. This, combined with the large
isospin-conserving $\pi\pi\rho$ coupling, could have a substantial
impact on the triangle discrepancy. However, as we shall see below,
in our model all isospin violations arising from the vector-meson
sector must vanish as $q^{2} \rightarrow 0$.

\section{Isospin-violating meson-nucleon coupling constants}
\label{seciv}

  In this section we concentrate on isospin violations to the
triangle identity which arise, exclusively, from isospin-violating
meson-nucleon coupling constants. Additional isospin-breaking
mechanisms, particularly those associated with Coulomb effects and
hadronic mass differences, have been treated elsewhere~\cite{gak95}.
Recently, we have estimated the effect of isospin-violating
meson-nucleon coupling constants on the $NN$ potential~\cite{ghp95}.
We have reported a large contribution from vector-meson exchange
to the class IV nucleon-nucleon potential. The isospin-violating
couplings that we have computed emerged from evaluating matrix
elements of quark currents between nucleon states; the violations
are driven by the up-down quark mass difference.

  The isospin violations that we have computed arise on
rather general grounds; we have assumed that the vector
mesons ($\omega$ and $\rho$) couple to appropriate isospin
components of the quark electromagnetic current. Moreover,
at $q^2=0$ our results are insensitive to the quark-momentum
distribution; they depend merely on the spin and flavor structure
of the nucleon wave function. As a result, some important
constraints emerge at $q^2=0$. In particular, only isospin
violations in the tensor (or anomalous) couplings are allowed
at $q^2=0$; the vector couplings are ``protected'' by gauge
invariance and remain unchanged. However, since all tensor-driven
contributions to $\pi N$ scattering vanish in the soft-pion limit
($q_{\mu} \rightarrow 0$) isospin-violating vector-meson-nucleon
coupling constants can not contribute to the triangle discrepancy.
Moreover, there is no contribution from $\rho$-$\omega$ mixing
at $q^2=0$~\cite{piewil93,oconn94}. Note that, contrary to the
claim of Ref.~\cite{cohmil95}, the momentum-dependence of the
$\rho$-$\omega$ mixing amplitude can not be absorbed into the
vertex without violating gauge invariance. Thus, in our model,
all three sources of isospin breaking in the vector-meson sector
must vanish at $q^2=0$. In our model, there is no isospin-violating
$NN\sigma$ coupling either; the $NN\sigma$ vertex, which has the
same nonrelativistic limit as the timelike component of the vector,
is also protected at $q^2=0$.

  However, there is no symmetry that protects the $NN\pi$ coupling
at $q^2=0$. We are interested in computing the coupling of the neutral
pion to the nucleon in a nonrelativistic quark model. At $q^{2}=0$ the
coupling is determined from the spin and flavor content of the nucleon
wave function. In contrast, the isospin-violating coupling of the nucleon
to the charged pions is sensitive to the quark momentum distribution and,
therefore, more uncertain~\cite{miller90}. It seems, however, that under
reasonable assumptions the quark model is able to generate isospin-violating
($NN\pi^{\pm}$) couplings of comparable strength as those obtained in
conventional hadronic treatments based on the neutron-proton mass
difference. Presumably, these effects have been included in
Ref.~\cite{gak95}.

The most general form for the on-shell $NN\pi^0$ vertex function
consistent with Lorentz covariance and parity invariance is given
by
\begin{equation}
     g_{\lower 2pt \hbox{$\scriptstyle NN\pi$}}
    \Lambda^{5}_{\lower 2pt \hbox{$\scriptstyle NN\pi$}} =
     g_{\lower 2pt \hbox{$\scriptstyle NN\pi$}}
    \Big[ g^{\pi}_N \gamma^{5} \Big]  \;.
 \label{vertexa}
\end{equation}
Here $g_{\lower 2pt \hbox{$\scriptstyle NN\pi$}}$ is the
isospin-conserving $NN\pi$ coupling constant known phenomenologically
from fits to $NN$ phase shifts and to the properties of the deuteron:
$g_{\lower 2pt \hbox{$\scriptstyle NN\pi$}}^{2}/4\pi=
14.21$~\cite{machl87,machl89}. The isospin-violating component
is assumed to emerge from evaluating matrix elements of a flavor odd,
pseudoscalar quark current between nucleon states, i.e.,
 \begin{equation}
  \langle N(p',s') |
    \left[ {1\over 5}\, \bar{u} \gamma^{5} u
         - {1\over 5}\, \bar{d} \gamma^{5} d \right]
  | N(p,s) \rangle =
  \bar{U}(p',s')
   \Lambda^{5}_{\lower 2pt \hbox{$\scriptstyle NN\pi$}}
  U(p,s) \;.
 \label{vertexb}
\end{equation}
Here $U(p,s)$ denotes an on-shell nucleon spinor of mass $M_{N}$,
momentum $p$ and spin $s$. Moreover, the constituent quarks are
assumed elementary as no quark form factors are introduced. The
coupling constants are computed at $q^2=0$ by examining the
nonrelativistic reduction of Eq.~(\ref{vertexb}); this is the
essence of the quark-pion model of Mitra and Ross~\cite{mitra67}.
In particular, in this limit the derivation closely resembles that
which is used in computing the nucleon magnetic moments~\cite{perkins82}.
We obtain,
\begin{mathletters}
 \begin{eqnarray}
   {g^{\pi}_p \over 2M_{p}} &=&
    {4 \over 3} \left({+1/5 \over 2m_{u}}\right) -
    {1 \over 3} \left({-1/5 \over 2m_{d}}\right) =
   +{4 \over 30 m_{u}} + {1 \over 30 m_{d}} \;, \\
   {g^{\pi}_n \over 2M_{n}} &=&
    {4 \over 3} \left({-1/5 \over 2m_{d}}\right) -
    {1 \over 3} \left({+1/5 \over 2m_{u}}\right) =
   -{4 \over 30 m_{d}} - {1 \over 30 m_{u}} \;,
 \end{eqnarray}
\end{mathletters}
where $m_{u}$ and $m_{d}$ are the up and down constituent quark
masses. Alternatively, one can construct nucleon isoscalar and
isovector combinations:
\begin{equation}
    {g^{\pi}_p \over 2M_{p}}\;{1 \over 2}(1+\tau_z) +
    {g^{\pi}_n \over 2M_{n}}\;{1 \over 2}(1-\tau_z) =
    {1 \over 6m} \left(
    {3\over 10}{\Delta m \over m} + \tau_z \right) \equiv
    {1 \over 2M} \Big(g^{\pi}_0  + g^{\pi}_1\tau_z\Big) \;.
\end{equation}
Note that we have introduced the following definitions:
\begin{equation}
    M \equiv {1 \over 2}(M_{n}+M_{p}) \;; \quad
    m \equiv {1 \over 2}(m_{d}+m_{u}) \;; \quad
    \Delta m \equiv (m_{d}-m_{u}) \;.
\end{equation}
The above relations are correct to leading order in $\Delta m /m$.
Moreover, they reveal an isospin-violating component ($g^{\pi}_0$)
in the $NN\pi^{0}$ coupling constant. In particular, by selecting
$m=M/3=313$~MeV and $\Delta m=4.1$~MeV~\cite{licht89} we obtain:
\begin{equation}
  g^{\pi}_0 = {3\over 10}{\Delta m \over m} \approx 0.004 \;.
\end{equation}
Ultimately, this isospin-violation can be traced back to the
up-down quark mass difference; the up quark, which is lighter,
generates a stronger coupling of the neutral pion to the proton
than to the neutron. Note that the isospin breaking computed in
the quark model is substantially larger---by about a factor of
six---than in the nucleon model of Ref.~\cite{cheung80} where
the scale of the breaking is set by the neutron-proton mass
difference. In contrast, for the coupling of the nucleon to
charged pions both models seem to generate an isospin violation
of comparable strength~\cite{miller90}.

  Incorporating the isospin-violating correction from $g^{\pi}_0$
into the evaluation of the triangle discrepancy is straightforward.
First, the elastic $\pi^{\pm}p$ amplitudes remain unchanged.
Second, it modifies the charge-exchange (CEX) amplitude
$f(\pi^{-}p \rightarrow \pi^{0}n)$ through a simple
renormalization of the nucleon Born terms; the s-channel,
which has a neutron in the intermediate state, gets reduced
relative to the u-channel, which contains a proton in
the intermediate state. Thus, in computing the charge-exchange
amplitude one must use an isospin-odd contribution given by
[see Eq.~(\ref{bminus})]:
\begin{equation}
    \widetilde{B}^{(-)}(s,t) \equiv
   -{g_{\lower 2pt \hbox{$\scriptstyle NN\pi$}}^{2}
    (1-g^{\pi}_0) \over s-M^{2}}
   -{g_{\lower 2pt \hbox{$\scriptstyle NN\pi$}}^{2}
    (1+g^{\pi}_0)\over u-M^{2}} =
     B^{(-)}(s,t) - g^{\pi}_0 \, B^{(+)}(s,t) \;.
 \label{btilde}
\end{equation}
Note that the ``small'' isospin-odd contribution $B^{(-)}$
is being corrected by the ``large'' isospin-even term $B^{(+)}$.
Indeed, at threshold $|B^{(+)}/B^{(-)}|=2M/m_{\pi} \approx 14$.
Now, however, there is no cancellation due to chiral symmetry;
there is no isospin-violating $NN\sigma$ coupling at $q^2=0$.
Using the above expression for $\widetilde{B}^{(-)}$ we
compute the value of the triangle discrepancy at threshold.
We obtain,
\begin{equation}
    D = - \sqrt{2} \;
    {g_{\lower 2pt \hbox{$\scriptstyle NN\pi$}}^{2}\over 4\pi}
    {g^{\pi}_0 \over M}
    {1 \over (1+m_{\pi}/M)(1-m_{\pi}^{2}/4M^{2})}
    \mathop{\longrightarrow}\limits_{m_{\pi} \rightarrow 0}
        - \sqrt{2} \;
    {g_{\lower 2pt \hbox{$\scriptstyle NN\pi$}}^{2}\over 4\pi}
    {g^{\pi}_0 \over M} \;.
\end{equation}
This generates an isospin violation to the triangle identity of
$D=-0.0145$~fm. The s-wave contribution to the triangle discrepancy
shows a very weak energy dependence. Indeed, its contribution at
$T_{\rm lab}=40$~MeV is $D=-0.014$~fm; we obtain a much smaller effect
from the p-waves: $1.3\times 10^{-4}$~fm and $-2.0\times 10^{-4}$~fm
for the $1^{+}$ and $1^{-}$ partial waves, respectively. This result
is in good agreement with the value reported recently by Gibbs, Ai,
and Kaufmann of $D=-0.012\pm 0.003$~fm from the s-wave alone or
$D=-0.011\pm 0.003$~fm for the sum of s and p waves at 40 MeV~\cite{gak95}

\section{Conclusions}
\label{secconcl}

  We have examined violations to the triangle identity that
arise from isospin-violating meson-nucleon coupling constants.
In our model, gauge invariance precludes the contribution from
vector-meson exchanges at $q^2=0$; these include $\rho$-$\omega$
mixing as well as isospin-violating $NN\omega$ and $NN\rho$
coupling constants. There is no symmetry, however, that protects
the $NN\pi^{0}$ coupling at threshold. We have computed isospin
violations in the $NN\pi^{0}$ coupling using a nonrelativistic
quark model. We have obtained a larger coupling of the neutral
pion to the proton than to the neutron as a result of the up
quark being lighter than the down quark. The observed isospin
violation is about a factor of six larger than the one computed
in nucleon models where the breaking is generated by the
neutron-proton mass difference. These results were used to modify
the relative weights of the s- and u-channel contributions to the
charge-exchange reaction $\pi^{-}p\rightarrow\pi^{0}n$.

  The isospin violation in the CEX amplitude became proportional
to the large isospin-even amplitude $B^{(+)}$; this amplitude does
not vanish in the chiral limit. In chiral models, such as the linear
$\sigma$ model used here, the large contribution from $B^{(+)}$
to the isospin-even scattering length is cancelled by an almost
equally large and opposite contribution [$A^{(+)}$] arising
from $\sigma$-meson exchange. However, in our model all isospin
violations in the $NN\sigma$ coupling must vanish at $q^{2}=0$.
As a result, we obtained a large violation to the triangle identity:
$D=-0.014$~fm. This value is in good agreement to the one reported
from a recent analysis of high-quality $\pi N$ data which yielded
$D=-0.012\pm 0.003$~fm~\cite{gak95}.

A particularly interesting test of this mechanism could be
a comparison of the ``mirror'' reactions
$\pi^{-}p\rightarrow\pi^{0}n$ and
$\pi^{+}n\rightarrow\pi^{0}p$~\cite{gak95}. For the first case,
namely, the one treated here, it was the s-channel that was
suppressed relative to the u-channel. In contrast, it is the
s-channel---now with a proton in the intermediate state---that
becomes enhanced in the $\pi^{+}n\rightarrow\pi^{0}p$ reaction.
One could quantify this isospin violation by measuring the difference
of these two amplitudes, i.e.,
\begin{equation}
   \widetilde{D} \equiv
     f(\pi^{-}p \rightarrow \pi^{0}n) -
     f(\pi^{+}n \rightarrow \pi^{0}p) \;.
 \label{triangleb}
\end{equation}
Note that the difference between the $pp\pi^{0}$ and $nn\pi^{0}$
coupling constants, alone, gives $\widetilde{D}=2D\approx-0.029$~fm.
This value should be compared to a charge-exchange scattering length
of $a_{0}=-0.19$~fm---it represents an isospin violation of 15\%.

Undoubtedly, much work remains to be done before a clear
understanding of the underlying mechanism behind the large
isospin violation reported in Ref.~\cite{gak95} will emerge.
Yet, we believe that isospin violations in the $NN\pi^{0}$
coupling constant are likely to play an important role in the
final analysis.

\acknowledgments
I thank S. Gardner, C.J. Horowitz, and B.D. Serot for many
helpful conversations. This work was supported by the DOE
under Contracts Nos. DE-FC05-85ER250000 and DE-FG05-92ER40750.

\end{document}